\documentclass[aps,pre,superscriptaddress,floatfix,11pt]{revtex4-1}
\pdfoutput=1
\usepackage{amsmath,amsfonts,amssymb}
\usepackage{graphicx}
\usepackage{epstopdf}
\usepackage{float,placeins}
\usepackage[hyperfigures,breaklinks,colorlinks,citecolor=blue,linkcolor=black]{hyperref}
\usepackage{subfigure}
\usepackage{mathbbol}
\DeclareMathOperator{\sign}{sign}

\begin{document}
\title{Low Auto-correlation Binary Sequences explored using Warning Propagation}

\author{I. Kotsireas} 
\affiliation{Wilfrid Laurier University. Waterloo, Canada}
\author{A. Lage-Castellanos} 
\affiliation{Group of Complex Systems and Statistical Physics. Department of Theoretical Physics, Physics Faculty, University of Havana, Cuba}
\email{mulet@fisica.uh.cu}
\author{O. E. Mart\'{\i}nez-Durive}
\affiliation{Group of Complex Systems and Statistical Physics. Department of Theoretical Physics, Physics Faculty, University of Havana, Cuba}
\email{orlando@estudiantes.matcom.uh.cu}
\author{R. Mulet}
\affiliation{Group of Complex Systems and Statistical Physics. Department of Theoretical Physics, Physics Faculty, University of Havana, Cuba}

\date{\today}

\begin{abstract}
  \noindent
  The search of binary sequences with low auto-correlations (LABS) is a discrete combinatorial optimization problem contained in the NP-hard computational complexity class. We study this problem using Warning Propagation (WP) , a message passing algorithm, and compare the performance of the algorithm in the original problem and in two different disordered versions. We show that in all the cases Warning Propagation converges to low energy minima of the solution space. Our results highlight the importance  of the local structure of the interaction graph of the variables for the convergence time of the algorithm and for the quality of the solutions obtained by WP. While in general the algorithm does not provide the optimal solutions in large systems it does provide, in polynomial time, solutions that are energetically similar to the optimal ones. Moreover, we designed hybrid models that interpolate between the standard LABS problem and the disordered versions of it, and exploit them to improved the convergence time of WP and the quality of the solutions.  
  
\end{abstract}

\maketitle

\section{Introduction}

The Low Auto-correlation Binary Sequence (LABS) problem consists in finding a binary sequence $ \{S\}=\{s_1, s_2,..., s_n\} $ where $ s_i \in \{1,-1\}  $ for $1\leq i \leq n$ that minimizes the function
\begin{equation}
E(\{S\})=\sum_{k=1}^{n-1}C_k(\{S\})^2 \label{eq:energy}
\end{equation}
\noindent where $C_k(\{S\})$ are the aperiodic auto-correlation coefficients: 
\begin{equation}
C_k(S)=\sum_{i=1}^{n-k}s_is_{i+k}. \label{eq:autocorr}
\end{equation}
\noindent LABS is an NP-hard optimization problem \cite{BoskovicBB17}.


To build these low auto-correlation binary sequences is of fundamental interest in many practical applications. In radars, for example, LABS are required for modulation in the process of pulse compression to enhance range resolution and long range detection capabilities \cite{SR14} and for the measurement of space-time curvatures between high precision radars \cite{Mertens16}. 
In Mathematics it is known as the Littewood problem, that consists in finding the coefficients of a polynomial around the unit circle in the complex plane. 
In Statistical Physics, these are the ground states of the Bernasconi's model \cite{Bernasconi87}, that implies the energy minimization of an Ising spin system with long range interacting variables with four-fold antiferromagnetic interactions.
Morover, it has find its way in digital signaling processing \cite{Kratica12}, 
and in Artificial Inteligence \cite{Amaya13}.

The problem has been largely studied using exact and heuristic methods. Exhaustive search is currently the only exact way to get LABS solutions. Golay published in \cite{Golay} optimal solutions for values up to $N=32$. Since then, due the computational complexity of the problem, all the new optimal solutions have been obtained using Branch and Bound \cite{Mertens16} methods with different parallel implementations, resulting in algorithms with cost of order $O(k^n )$ with $k < 2$.

For example, Mertens found the solution of the LABS problem up to $n \leq 48$ with a computational cost of
$O(1.85^n)$ in 1996 \cite{Mertens96}. 
Later in 2004 with an improved implementation Mertens and Bauke computed the solution up to $n \leq 60$ \cite{Mertens16}. Wiggenbrock expanded this approach to $n \leq 64$ using a better bound and reported a cost of order $O(1.79^n)$ \cite{Mertens16}. 
More recently in 2016, Packebusch and Mertens obtained two more values $n = 65$ and $n = 66$, employing the combined bounds of Prestwich and Wiggenbrock $O(1.729^n)$ \cite{Mertens16}. 
Despite these sophisticated implementations and improvements it is clear that this approach is not viable in the search of optimal sequences with larger lengths, for example, $n > 100$.

A readable summary of the application of different heuristic and stochastic algorithms to LABS appears in  \cite{Amaya13} and \cite{BoskovicBB17}. 
More specifically, in \cite{Militzer} the authors used  Evolutionary Search (ES) taking a special care on processes to generate the offsprings without a recombination strategy and with a mutation operator flipping more than one bit at once, the authors found solutions, for odd system sizes, as large as $N=201$. 
In \cite{Gallardo} the authors presents empirical evidence showing that evolutionary pure algorithms are not capable of facing the complexity of the problem, but also that evolutionary methods assisted by local-search operators (memetic algorithms) provide optimal or near-optimal results. 
In \cite{Halim} the authors improved the Tabu Search Method previously developed in \cite{Dotu} getting better results than memetic algorithms and they explore the solution space for systems size between $[61-77]$ and report the best values that they obtain with TSv7 algorithm which has an expected running time of $\mathcal{O}(1.03e-5*1.34^n)$, according to these authors the improvement came from a deep study of the searched trajectories and a better diversification process.
The state of the art algorithms in this heuristic field according to \cite{Mertens16} are the work  \cite{BoskovicBB17} where the authors combined a random self-avoiding walk and Hasse graph. 
However, like with exact algorithms these complex and advanced heuristics fail in systems of relative high dimensions \cite{Mertens16} leaving a lot of room for further improvements.

On the other hand, already in 1994 a series of papers followed a different path \cite{Bouchaud94, Marinari94, Migliorini94}. 
They approached the problem  borrowing techniques and concepts from the statistical physics of disordered systems, and the model, since then, became archetipical of the existance of glassy phases in models without disorder. The basic idea then was to study a disordered version of the LABS problem and to predict its average properties in the infinite size limit. 
In short, one must notice that it is possible to write equation \eqref{eq:energy} as:
\begin{equation}
  E(\{S\})=\sum_{k=1}^{n-1}(\sum_{i,j}J^k_{i,j} s_i s_j)^2  \label{eq:energyWithMatrixs}
\end{equation}
\noindent where for LABS $J^k_{i,j}$ is defined as:
\begin{equation}
        J^k_{i,j} = 
        \left\{
        \begin{array}{c l}
        1 & j = i+k \\
        0 & \mbox{otherwise}
        \end{array}
        \right. 
        \label{Labs-Matrix}
\end{equation}
\noindent while a disordered version of the model \cite{Bouchaud94} may look like $J^k_{i,j}$:
\begin{equation}
        J^k_{i,j} = 
        \left\{
        \begin{array}{c l}
        1 & \mbox{ with probability } (n-k)/n^2 \\
        0 & \mbox{otherwise}
        \end{array}
        \right. 
        \label{Bouchaud-Matrix}
\end{equation}
\noindent This is a sort of Mean Field (MF) version of the problem that preserves the connectivity of the variables, but diminishes the correlations between them randomizing $J^k_{i,j}$. 
Studying such a model it was possible to introduce ideas and methods developed for spin glasses \cite{MPV87} and to find that the system undergoes a first order transition with a glassy phase at low temperatures, much as if quenched disorder were present. 
Another model, which reminds the original LABS problem, is the anti-ferromagnetic PSpin model with (p=4) defined by the energy function:
\begin{equation}
E(\{S\}) = \sum_{i,j,k,l} J_{i,j,k,l} s_i s_j s_k s_l
\label{eq:ener-pspin}
\end{equation}
\noindent where $J_{i,j,k,l}$ is a random diluted matrix with elements $0$ and $1$, again with the same number of interactions than the LABS model. 
Although the model is well understood in the diluted and fully connected regimes \cite{dilutedricci,SGTFP}, as far as we know it has never been compared with LABS or its mean field version.

The three models share a similar formal structure, defined by the interaction of 4 variables. 
However, while in the original LABS problem the variables are strongly correlated, this correlation diminishes in the MF model by the randomicity of $J_{ij}$, but still is not zero because of the square in \eqref{eq:energyWithMatrixs}. 
On the other hand, the PSpin model defined by \eqref{eq:ener-pspin} can be viewed as a completely disordered Hamiltonian where the connections between the variables are absolutely random. 
Therefore, a comparison between the three models may be helpful to elucidate the role of the correlations in the graph structure on the spectra of the Hamiltonian and on the performance of the algorithms.

Independently, since the beginning of the century,  Message Passing algorithms have found their way in the realm of the statistical physics community. 
The turning point was the realization that message passing algorithms can be viewed as fixed point equations derived from variational approximations to the free energy of Ising-like models \cite{Yedidia02}. 
This inspired researchers to look for novel applications of these algorithms in the field of Combinatorial Optimization \cite{Mulet02, Braunstein03, Weigt06} and to explore new extensions \cite{SP,CVM,GBP}. 

In this context our work studies the performance of Warning Propagation, in the solution of the LABS problem and its disordered versions, motivated here as a simple heuristics that is well understood \cite{Weigt06}. 
We first shed light on the relevance of the structure of the underlying graph of interactions on the space of available configurations of the problem. 
Then, we study the performance of the algorithm and how it depends on the structure of the graph and  show that although for relatively large system sizes the solutions found for LABS are not the optimal ones, they nevertheless exhibit very low energies. 

\section{Warning Propagation}
\label{sec:wp}

LABS problem can be represented by a standard factor graph, where both the variables and their interactions are graphically represented. Hamiltonians (cost functions) of the types (\ref{eq:energy}), (\ref{eq:energyWithMatrixs}) and (\ref{eq:ener-pspin}) are all made of the sum of groups of sizes 4 and 2 interacting variables. 
Let us take as an example the part $C_1(S)^2$ of the LABS hamiltonian  (\ref{eq:energy}) with $N=5$ spins:

\begin{eqnarray}
 C_1(S)^2 &=& \left( \sum_{i=1}^{5-1} s_i s_{i+1} \right)^2 = (s_1 s_2 + s_2 s_3 +s_3 s_4 +s_4 s_5)^2 \nonumber \\
&=& \mbox{const}+ 2 s_1 s_3 + 2 s_2 s_4 + 2 s_3 s_5 + 2 s_1 s_2 s_3 s_4 + 2 s_1 s_2 s_4 s_5 + 2 s_2 s_3 s_4 s_5 \label{eq:C1} 
\end{eqnarray}

\begin{figure}[h]
\centering
\includegraphics[scale=0.4]{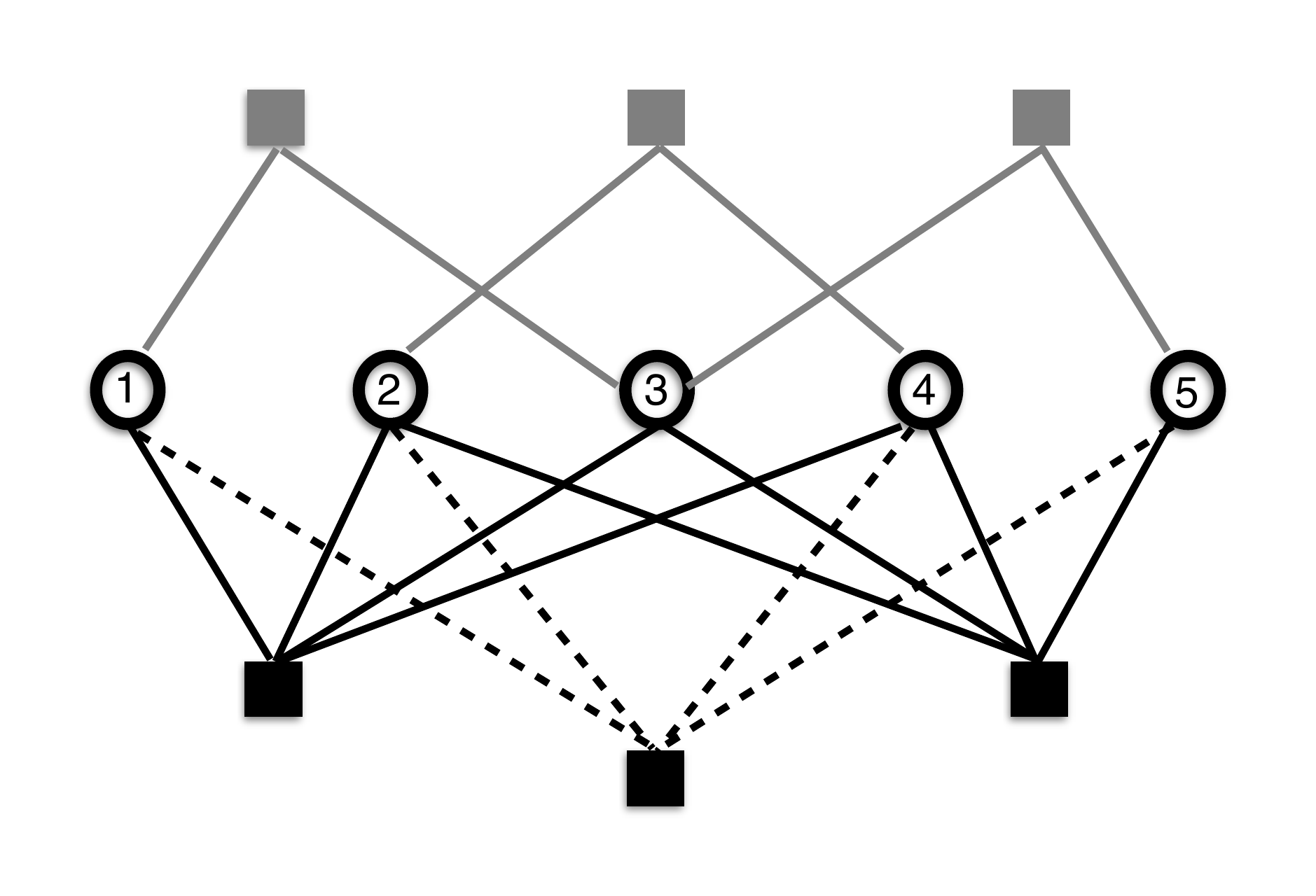}
\caption{Graphical representation of quation (\ref{eq:C1}). The variables (cirlces) are connected to interactions (square nodes). In gray, the three first terms (connecting only two variables), in black the other three 4-spins interactions.} 
\label{fig:factorgraph}
\end{figure}

\noindent We have used the fact that binary spins $s_i^2 \equiv 1$, so every square value can be disregarded as constant.
Each summand coorrespond to an interaction, represented in the factor graph as a square node (see Fig. \ref{fig:factorgraph}), while the variables interacting are represented as circles, joined to their corresponding interactions by an edge in the factor graph. Later we will use $a,b,c$ letters to refer to the indexes of the factor nodes (interactions), and $i,j,k$ to refer to the indexes of the variables.

Although message passing algorithms, like sum-product \cite{kschischang2001factor}, belief propagation  \cite{Yedidia02}, and warning propagation \cite{ZPhD08} have been derived more than once in different communities and for different application, with different balances between rigor and intuition, we (physicist and computer scientics) prefer to think of them as an approximation to study the properties of the measure:

\begin{equation}
 P(S) = \frac 1 Z \exp\left(-\beta E(S) \right).
\end{equation}

\noindent In the limit of low temperatures ($\beta  = 1/T \to \infty$) the measure concentrates on the configurations of lowest energy, and therefore good approximations can be transformed into good optimization procedures. 
Warning propagation corresponds to the zero temperature limit of the Bethe approximation in statistical mechanics. 
Said in other terms, the max-sum algorithm corresponds to the zero termperature limit of the sum-product one (standard belief propagation).

We will implement WP in terms of two types of messages: $U_{a \to i}$ carrying information from interaction node $a$ to any of its variables $i$, and the converse $H_{i\to a}$ sending information from variable $i$ to one of its interactions. You can derive the following two equations relating these messages by 
taking the appropriate $T=0$ limit in the belief propagation equations:
\begin{eqnarray}
U_{a \to i} &=& -\sign{(H_{j \to a}H_{l \to a}H_{k \to a})} = -\sign\left( \prod_{j\in a\setminus i } H_{j\to a}  \right)  \label{eq:UMessage} \\
H_{i \to a} &=& \sum_{b \in N(i)\setminus a} U_{b \to i} \label{eq:HMessage}
\end{eqnarray}
The first equation is expanded for the case of factor nodes $a$ with four variables, but it should be clear from the compacted form that in the case of two-variables factors, the equation reduces to $U_{a \to i} = -\sign{(H_{j \to a})}$.

We can grasp some intuition on the meaning of these equations by looking at their similarity with the condition for a configuration to be a local energy minima of LABS:
\begin{eqnarray*}
U_{a \to i} &=& -\sign{(S_{j }S_{l}S_{k })}  \\
S_{i } &=& -\sign \left(\sum_{b \in N(i)} U_{b \to i } \right) 
\end{eqnarray*}
In this notation, $U_{a\to i}$ acts as the local opinion of the interaction $a$ on which direction should the variable $S_i$ be pointing to, while the second line ensures that each variable points to the direction that minimizes the value of the interactions with the majority of its factor nodes. 
To make even clearer the connection, we can write the minimum energy conditions as:
\begin{eqnarray*}
U_{a \to i} &=& -\sign{(H_{j }H_{l }H_{k })}  \\
H_{i } &=& \sum_{b \in N(i)} U_{b \to i }  \\
S_i &=& -\sign{ H_i }
\end{eqnarray*}
The WP equations are a ``cavity'' version of these conditions, where the cavity term refers to the fact that self interaction $U_{a\to i}$ is removed from (\ref{eq:HMessage}) in the definition of the field $H_{i \to a}$.

The Warning Propagation algorithm consists of iterating equations (\ref{eq:UMessage}) and (\ref{eq:HMessage}) until convergence. 
This kind of message passing algorithm are not guaranteed to converge, and there are many cases where they don't.
Luckily enough, in our case we found WP to be convergent almost every time, in spite of the fact that our topology of interactions (the factor graph) has many short loops. 
Once the equations are at a fixed point, we can recover a local energy minimum by setting the total field to $H_{i } = \sum_{b \in N(i)} U_{b \to i }$ and the spins to $S_i = -\sign{ H_{i }}$.

\section{Results}\label{sec:results}

\begin{figure}[hbtp]
\centering
\includegraphics[scale=0.7]{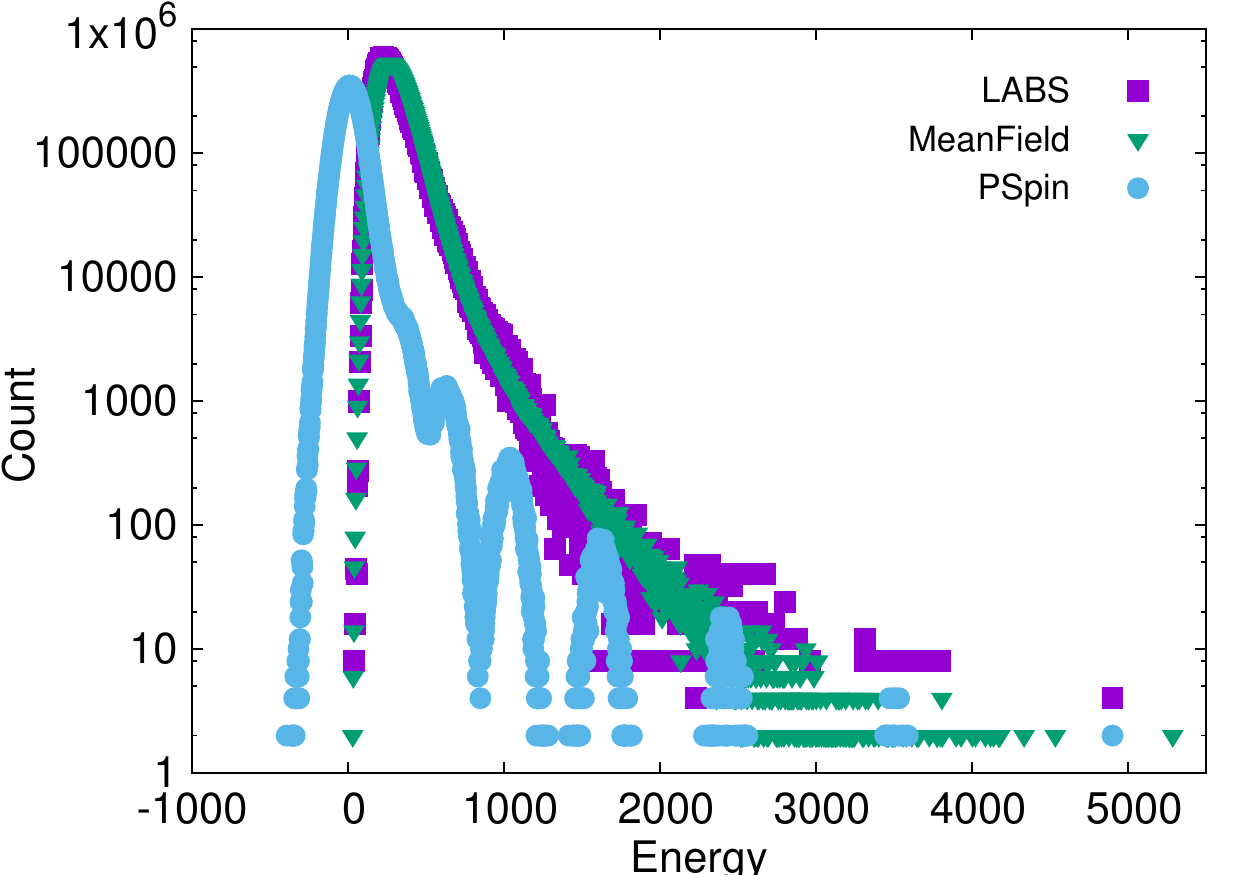} 
\caption{Histogram of energy of the configurations in the three models: LABS, Mean Field and PSpin. The system size is $N=25$ and the disorder models are self-averaging. There is  very similar behaviour between LASB and Mean Field model, the PSpin model exhibit a more rich structure for high energy configurations and allow configurations with negative energy value.}\label{histogram}
\end{figure}

We start the comparison between the three models, the original LABS, the Mean Field (MF) model and the $p$-spin (with $p=4$) antiferromagnet, plotting their corresponding histograms of the energies of all the possible configurations, extracted from exhaustive enumeration, for a given system size. 
The results are summarized in Figure \ref{histogram} where it becomes evident that while for LABS and MF the histograms are very similar the histogram of the PSpin model starts at lower energies and has a richer structure at high energies.

\begin{figure}[h]
\centering
\subfigure[LABS]{\includegraphics[scale=0.6]{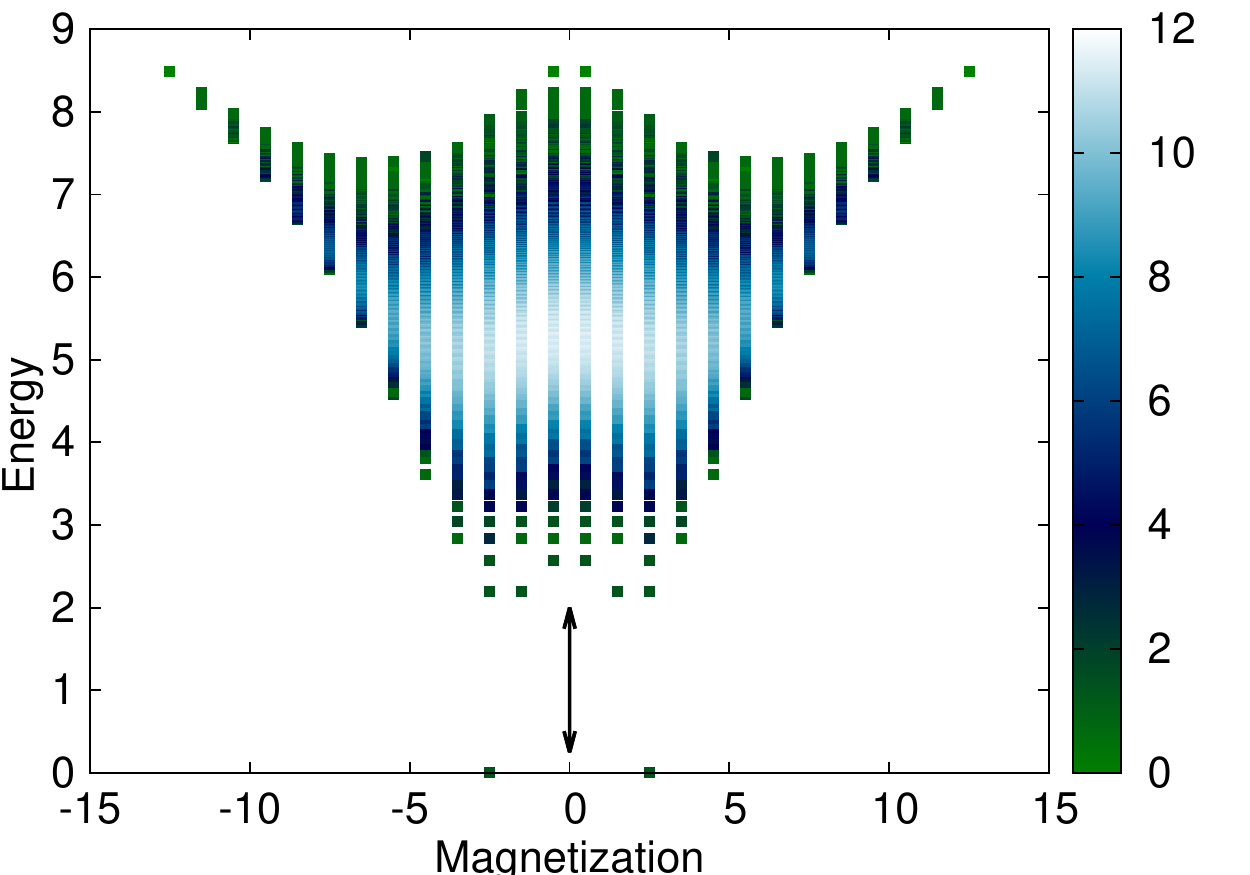}}
\subfigure[Mean Field]{\includegraphics[scale=0.6]{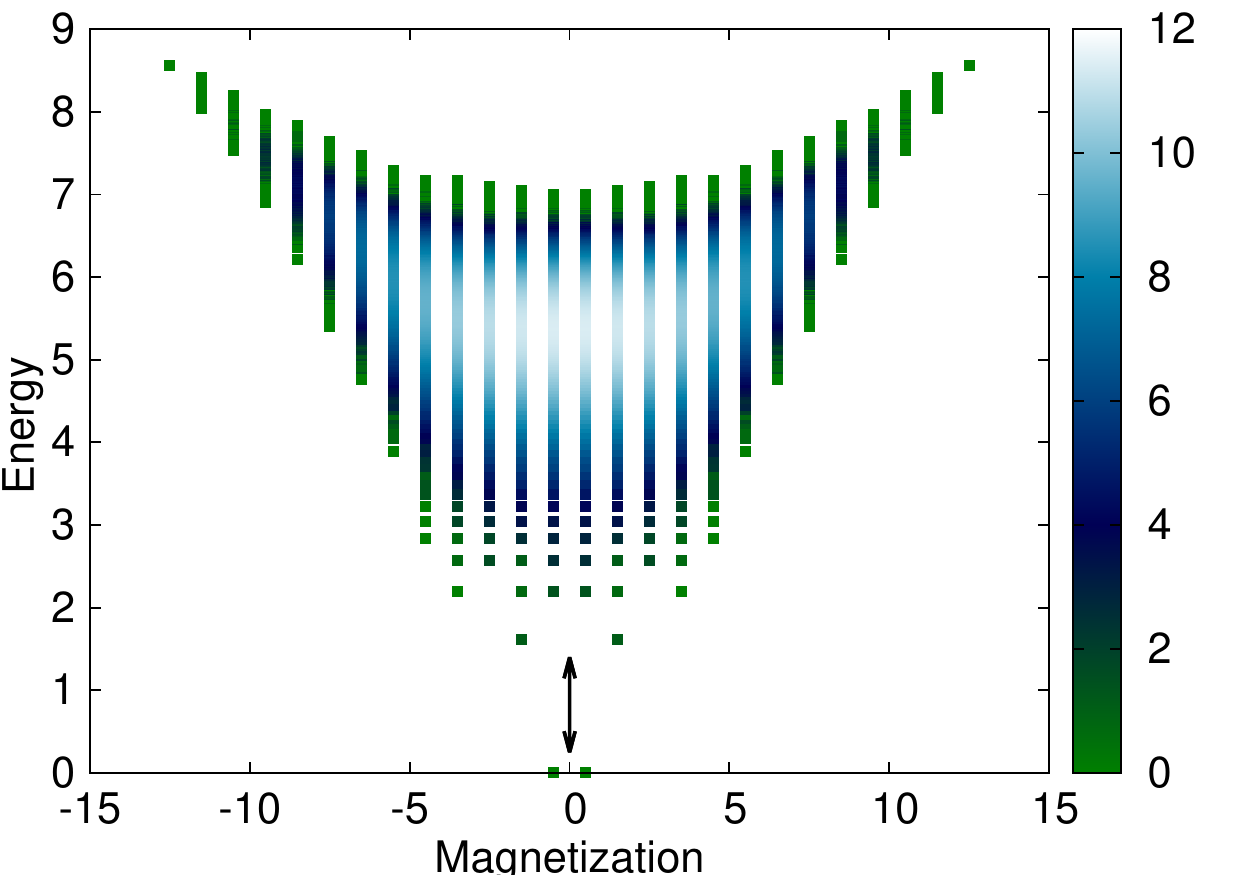}}
\subfigure[PSpin]{\includegraphics[scale=0.6]{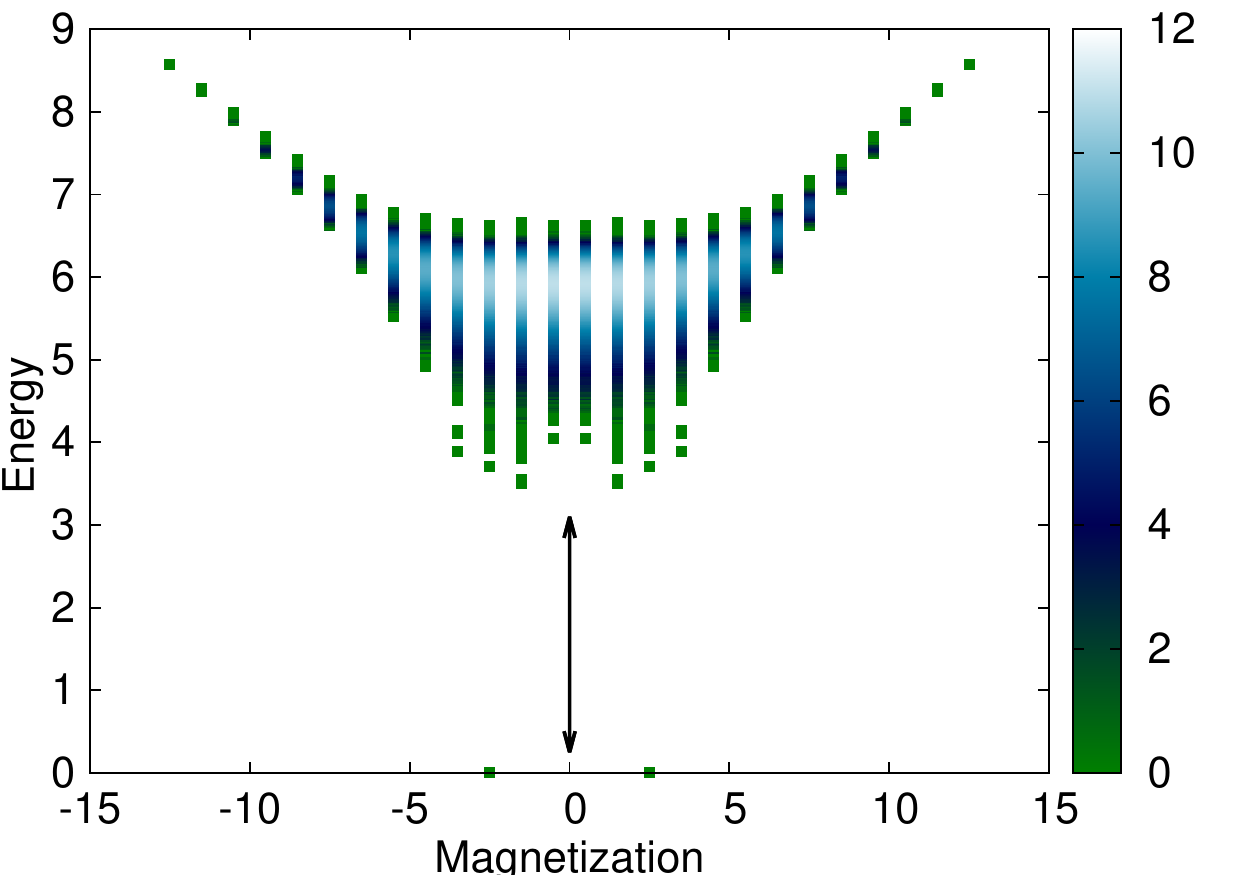}}
\caption{Structure of the solutions space for the three models: LABS, Mean Field and PSpin with three dimensions: energy value with a shift for start at zero in the $y$ axis in logarithm scale; the magnetization value in $x$ axis and finally the  number of configuration for a given energy and magnetization value in the color representation also in logarithm scale. Those systems have dimension $N=25$ and the disorder models are autoaverage.} 
\label{magStructure}
\end{figure}

A more detailed comparison appears in Fig. \ref{magStructure} where we show for the three models a bi-dimensional plot representing the number of configurations for a given energy and magnetization in systems of size $N=25$.
Despite the clear differences between the three panels, some general properties are very similar. 
Configurations with large energies have large magnetizations, indeed the ground state solutions are concentrated at low values of the magnetization. 
More importantly, there is a well defined gap between the three ground states of the models and the nearest configuration of lower energy.

\begin{figure}[hbtp]
\centering
\includegraphics[scale=0.7]{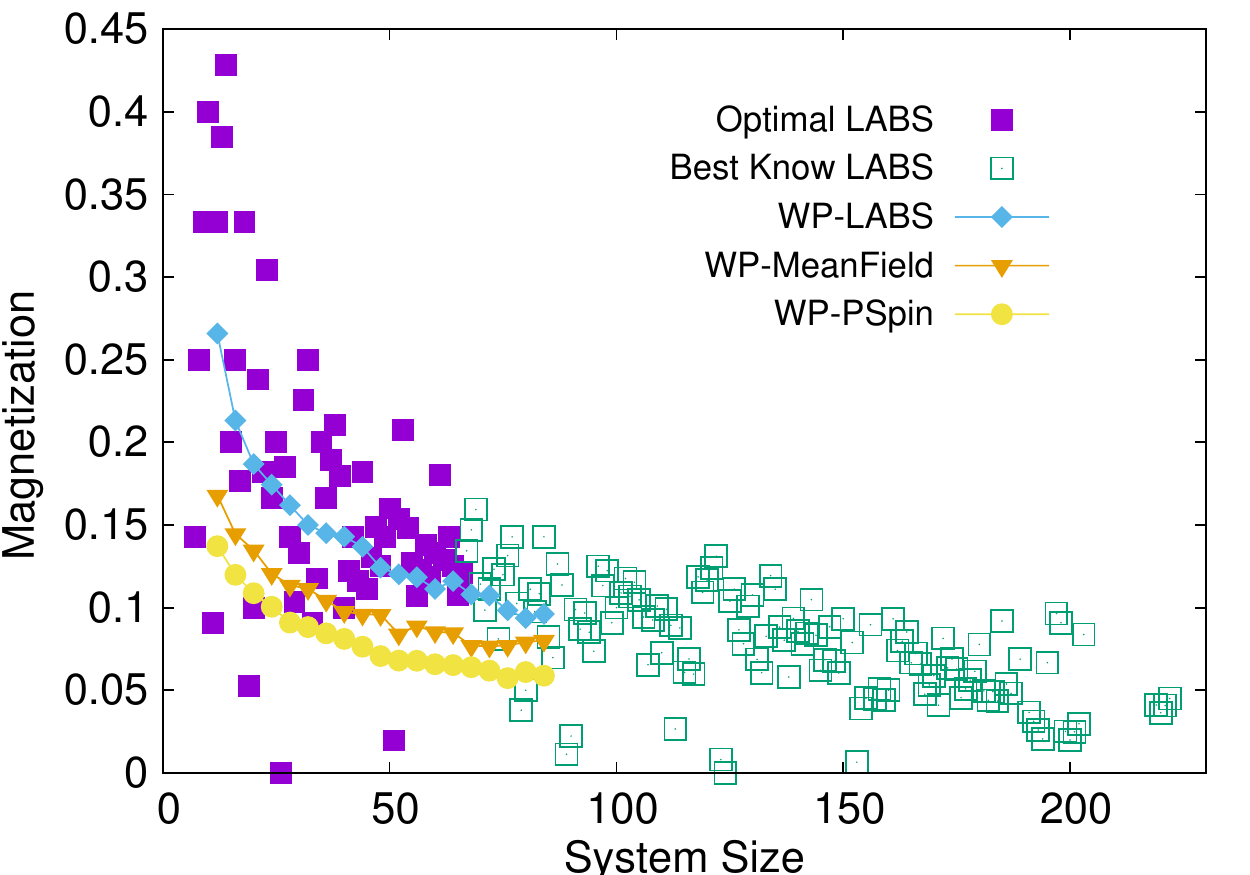} 
\caption{Magnetization of optimal and best know values for LABS problem in function of system size represent by the filled and unfilled square. 
The first, second and thrid continous lines are the average of the modulos of the magnetization of WP solutions for LABS, MeanField and PSpin models respectively, this points was obtained over a 1000 samples, this lines are very closed to the expected values for local minimal configuration of those systems.}\label{mag}
\end{figure}

We further explore the magnetization of the global optima (black squares in Figure \ref{mag}) and the best known values at large system sizes (open squares in Figure \ref{mag}). 
From the figure we may conclude that despite the strong fluctuations for low values of $N$ there is a clear trend suggesting that for large system sizes the global optimal are cofigurations of very low magnetizations, as one would guess for an atiferromagnetic model. 
Intuitively, a configuration with a large number of variables pointing in a prefered direction will produce many positive terms in \eqref{eq:energyWithMatrixs}. 
Therefore, low energy configurations should have a balance between the number of positive and negative spins. 
In the same figure, and for comparison, we plot the mean magnetization obtained after averaging over the different fixed point solutions of WP for different system sizes. 
The three models show the same dependency, the larger the system size $N$, the lower the magnetization. 
Moreover, the average magnetization of the fixed points of WP run for the LABS are consistently larger than the corresponding mean for the MF and PSpin models.

\begin{figure}[hbtp]
\centering
\includegraphics[scale=0.7]{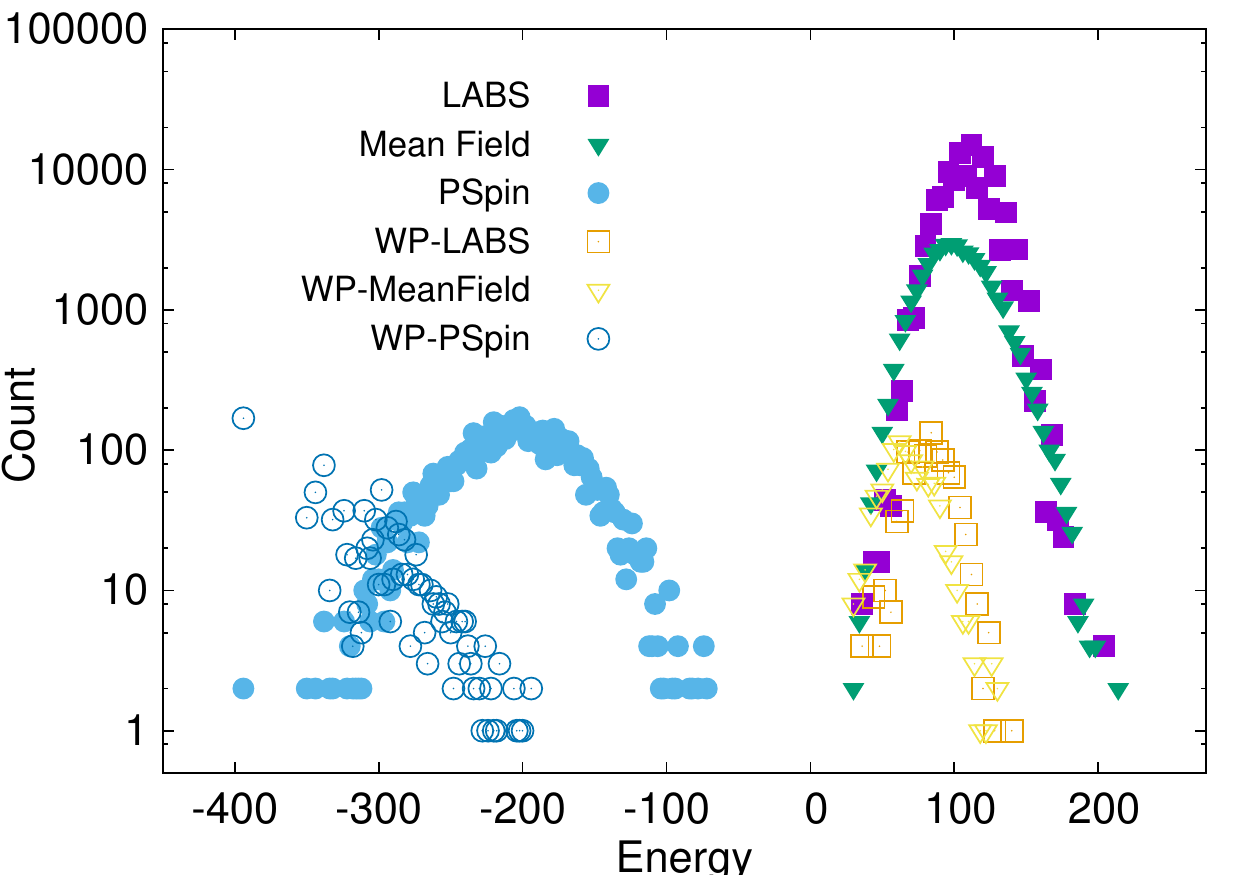}
\caption{Local minimum values for the three models: LABS, Mean Field and PSpin. Exhautive enumeration are represent by a filled symbol. The system size is $N=25$ and the disorder models are autoaverage. The unfilled symbols  represent the WP solutions for each model over $1000$ running. Is important to note that all the solution reach by WP in all models are local minimum configuration of those system.} \label{localminimum-wp}
\end{figure}

Already with a picture of the differences between the configurartions of the three models, and inspired by the success of WP reproducing the mean magnetization of the optimal solutions for the LABS problem we proceed focusing our attention on the local minima of the different problems. 
We define as a local minima a configuration whose energy can not change by a single spin-flip. 
For a system size $N=25$ we explore all the local minima in the three models by exhaustive enumeration and show in Figure \ref{localminimum-wp} the histograms of their energies (filled symbols). 
As for the global optima, one must notice that the structure of the histograms of the LABS problem and the MF model are very similar and clearly different from the histogram of the PSpin model. 
Then, we look for the fixed point solutions of WP for the different problems and plot the historgrams of the corresponding energies (open symbols in Figure \ref{localminimum-wp}). 
In the three models the solution of Warning Propagation concentrates in the low energy region of the local minima.
 This is already a point in favour of Warning Propagation as a good proxy to obtain low energy solutions for LABS.
More generally, it also suggests that more sophisticated message passing algorithms should be explored in this context.

Having described the main properties of the fixed points of Warning Propagation it is important to understand its convergence properties. 
To do this we design a series of models to interporlate between the disordered MF and PSpin model and the original LABS. 
In both scenarious we introduce a parameter $q \in [0,1]$  that fixes the order of the model. 
For the Mean Field model family we build the interaction between two variables inside the autocorrelations chosing with $q$ probability the expresion \eqref{Bouchaud-Matrix} and with probability $(1-q)$ the original expresion of LABS matrix \eqref{Labs-Matrix}.
In the PSpin version the idea is the same, but we use the random 4-spin interaction instead of the well correlated four variables choosing with $q$ probability terms in the expresion \eqref{eq:ener-pspin} and with $(1-q)$ probability term of \eqref{eq:energy}. 
For $q=0$ the disorder models are recovered and for $q=1$ the LABS original model. Then, we study the time for convergency of WP as a function of $q$.

\begin{figure}[h]
\centering
\subfigure[Model and Time]{\includegraphics[scale=0.6]{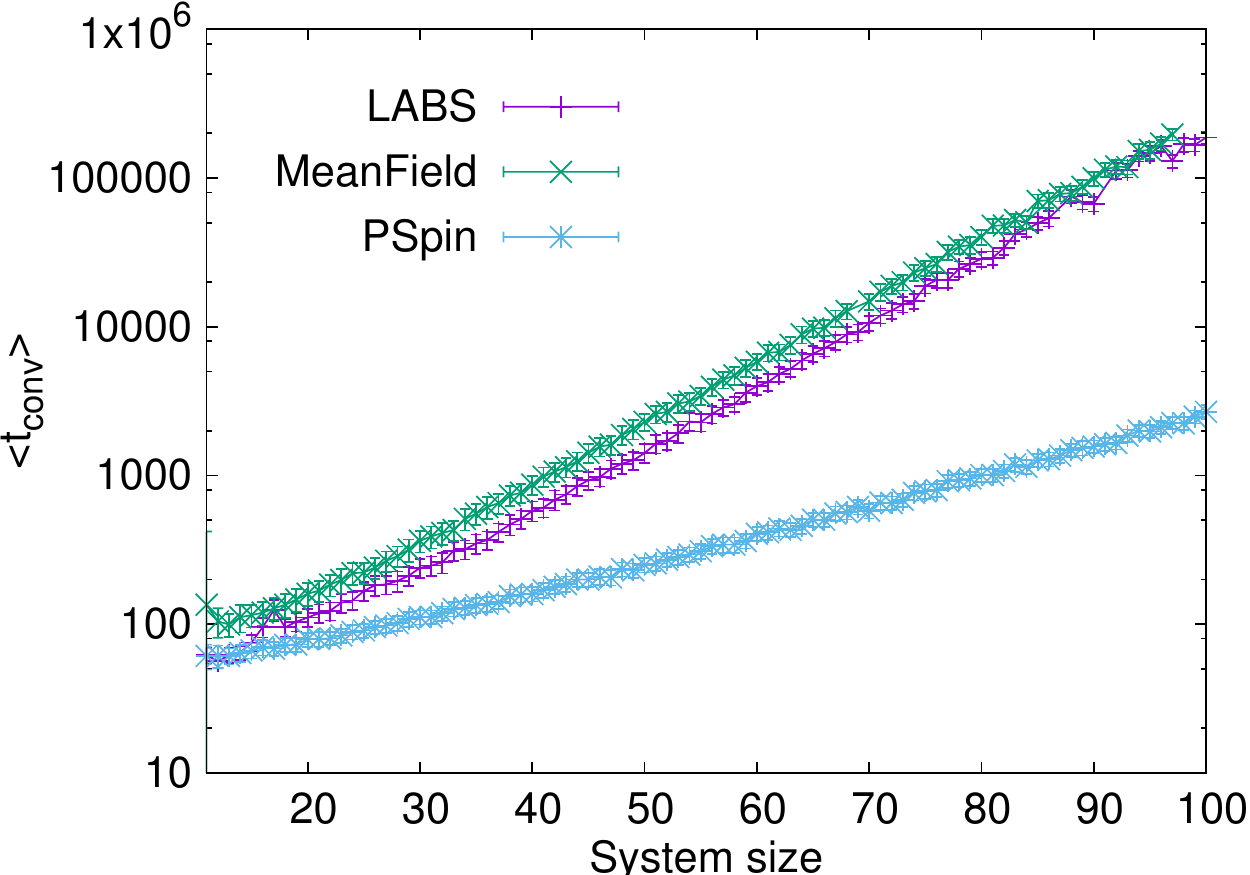} \label{modelTime}}
\subfigure[Transition Mean Field]{\includegraphics[scale=0.6]{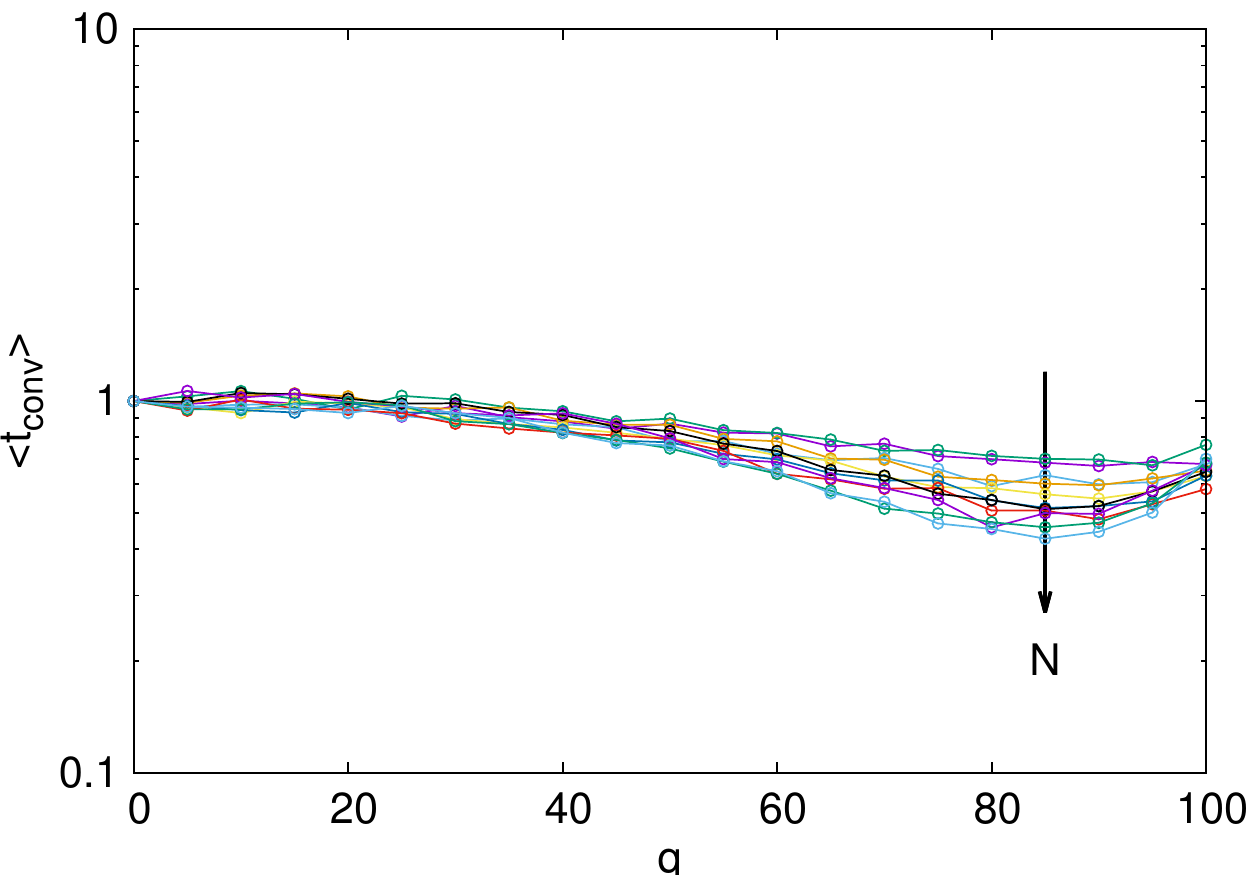} \label{MFTransition}}
\subfigure[Transition PSpin]{\includegraphics[scale=0.6]{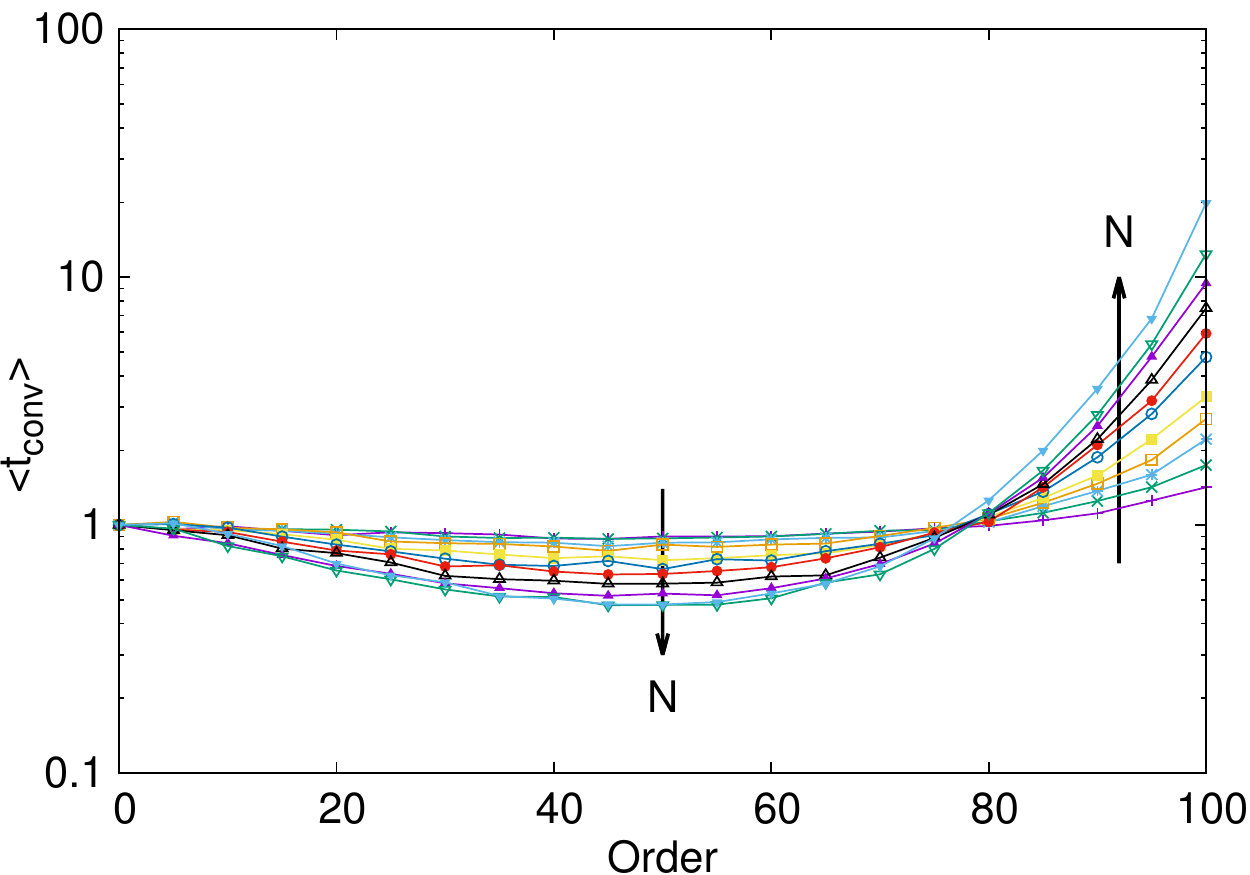} \label{PSpinTransition}}
\caption{
Behaviour in running time of WP algorithm over differents $q$ order parameters, building hibrid models: Mean Field and LABS in subfig a) and PSpin and LABS in subfig b). The $<t_{conv}>$ axis represent the necessary number of iteration to reach convergence normalized to the value of $q=0$ or "pure disorder model". In subfig c) we show the grow of the iteration number for each model where $q=0$ for Mean Field and PSpin.} 
\label{IterationTransition}
\end{figure}

In Fig. \ref{modelTime} we first present the convergence time for the three models as a function of $N$. While for the MF model and LABS the curves are very similar, it becomes evident that the convergence for the PSpin model is much faster. 
For $N=100$ at least two order of magnitude faster. 
However, the dependency of the convergence time for the models as a function of $q$ is less evident.

In  Fig. \ref{MFTransition} we show for different values of $N$ the time of convergence of different models (i.e. for different values of $q$ normalized with respect to the original Mean Field model). First one recalls from Fig. \ref{modelTime} that the convergence time of the MF and the original LABS problem, ($q=0$ and $q=1$), are quite similar. This is clearly reproduced in Fig. \ref{MFTransition}
too.  However, there is a specific value of $q \sim 0.8$ for which the convergence time is consistently shorter that in both original models, more than that, this difference increases with the system size.

We show in \ref{PSpinTransition} a similar plot but using the PSpin model as a reference. Again, we
 find a specific value of $q$ around 0.8 in which all curves cross at a time that is exactly the time expent by the original PSpin model to converge. Notice however, that in this case the convergence time is much lower than the usual convergence time of the original LABS (see subfig. \ref{modelTime}), in other words randomizing the $20\%$ of the interactions in the LABS problem, one obtain a resulting model that is very easy to solve.
It comes immediately out the following question: Could we exploit this fact to improve the convergence time of the algorithm for the original LABS problem? 

\begin{figure}[hbtp]
\centering
\includegraphics[scale=0.7]{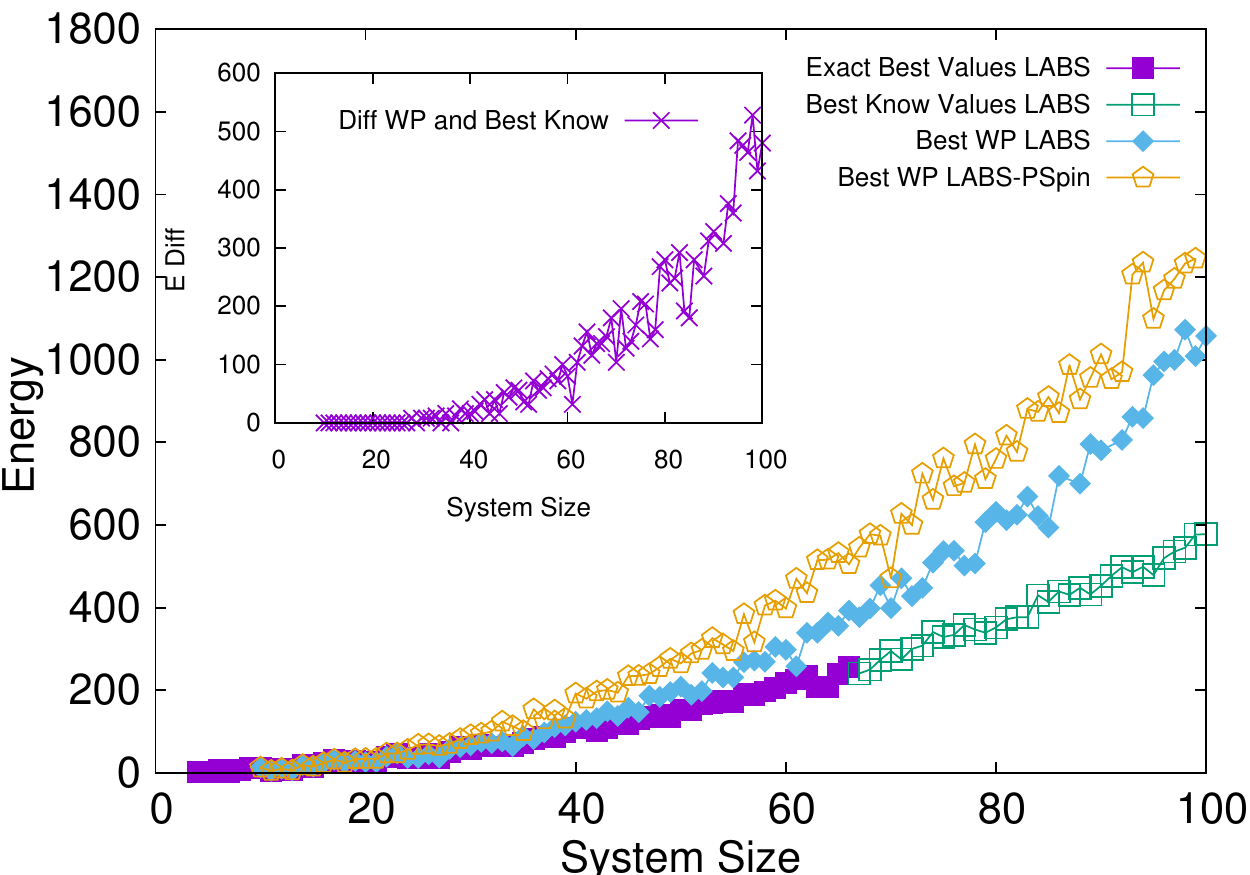} 
\caption{ General performance of WP over the LABS represent by the diamond curve and hibrid PSpin and LABS models at $q=0.8$ through the unfilled circle curve in the search of solutions. The filled square represent the optimal LABS solutions and the unfilled square the best know values. In the subplot we compare the best WP results with the best values.
Each point of WP was compute over a $1000$ samples.}
\label{performace}
\end{figure}

With the last question in mind we want to understand how good is Warning Propagation as an optimization heuristic for this kind of problem. 
In Fig \ref{performace} we plot together, the best known values for LABSP and the best results obtained running WP on the LABS and in the hybrid PSPin model with $q=0.8$. For relative small systems $n \leqslant 35$ WP run in LABSP recovers the true ground states of the system, but as the system size growths it get trapped in local minimal configurations supporting the idea of a complex energy landscape. Not surprisingly, the best known solution of WP run over the hybrid PSpin provides energies that are higher than the expected for WP run in the proper original LABS problem.
It is interesting to notice for example, that in the extreme case analyzed $N=100$, the best known value for the LABS problem is  $E=578$, while the best solution of WP we can report is $E=1058$ which is two times larger. However, one can not already discard it as a too high value. In fact, there exists a very pronounce energy gap between the value of the energy level of a global optimal and the second lowest energy level corresponding for some local optimal, evidence of this behaviour is show in Fig \ref{magStructure}. Moreover, the average energy defined as in LABSP of a random sequence of size $N=100$ is of order $5000$, i.e., we get a sequence that is five time better that a random sequence, and just two times worst that the best known solution to the LABS problem.

We can, however, go back to our question above and check if we can use the velocity of WP on hybrid instances of the PSpin model to improve the results of the WP over the original LABSP.  Using the same computer architecture, making a single observation,  for $M=100$ over LABS consume approximately two hours, meanwhile it's only necesary a quarter of hour to make a single observation in the hybrid model.
Then, we test, again for $M=100$, the performance of WP over the hybrid PSpin model at $q=0.8$, but, using the advantage of the faster convergence exploring more initial conditions, in this case up to $M=10000$ (If N was used to define the size of the system, can not be used to define the number of convergence attempts or intial conditions.). After convergency, these solutions were introduced again to WP, but, now run on the original LABSP problem. The convergence time was always of a few iteration steps. The final outcome was $E=914$, that although not yet the best known value, clearly better that the result obtained running directly WP on the original LABS problem.
\section{Acknowledgments}
This work was made possible by the facilities of the Shared Hierarchical Academic Research Computing Network (SHARCNET, http://www.sharcnet.ca) and Compute/Calcul Canada.
\section{Conclusions}

In this work we explored the relevance of the ordered structure in the LABS problem looking into the performance of a message passing algorithm on the original problem and in different disordered versions of it. We showed that it is possible to obtain sequences with low autocorrelation values using a very simple expression in a message passing framework defined by the Warning Propagation algorithm. The differences between our estimations and the best known values growth quadratically with the system size. This despite the fact that we use the more naive version of Warning Propagation and do not implement any optimization strategy, like parallel programing, decimation or especial initials conditions. Moreover, we showed that while the performance of WP in the mean field version of the problem is as slow as in LABSP, the performance in the PSpin version of the problem is much faster, supporting the idea that indeed these correlations are the fundamental cause for the difficulty of this problem. Based on this suggestion and the optimal performance of WP on a partially disordered version of LABSP, we showed that it is possible to exploit the velocity of WP to obtain sequence with very lower autocorrelation values. 


\bibliographystyle{plain}
\bibliography{bibliolabsv4}

\begin{thebibliography}{10}

\bibitem{Amaya13}
John~Edgar Amaya and Mar{\'i}a de los~{\'A}ngeles Tarazona.
\newblock Desempeño de metaheurísticas sin memoria en el problema del labs.
\newblock In {\em 11 th Latin American and Caribbean Conference for Engineering
  and Technology, August 14-16}, 2013.

\bibitem{Bernasconi87}
J.~Bernasconi.
\newblock Low autocorrelation binary sequences: statistical mechanics and
  configuration space analysis.
\newblock {\em J. Physique}, 48559, Oct 1987.

\bibitem{BoskovicBB17}
Borko Boskovic, Franc Brglez, and Janez Brest.
\newblock Low-autocorrelation binary sequences: On improved merit factors and
  runtime predictions to achieve them.
\newblock {\em Appl. Soft Comput.}, 56:262--285, 2017.

\bibitem{Bouchaud94}
Jean-Philippe Bouchaud and MARC M{\'e}zard.
\newblock Self induced quenched disorder: a model for the glass transition.
\newblock {\em Journal de Physique I}, 4(8):1109--1114, 1994.

\bibitem{Braunstein03}
A.~Braunstein, R.~Mulet, A.~Pagnani, M.~Weigt, and R.~Zecchina.
\newblock Polynomial iterative algorithms for coloring and analyzing random
  graphs.
\newblock {\em Phys. Rev. E}, 68:036702, Sep 2003.

\bibitem{SGTFP}
Tommaso Castellani and Andrea Cavagna.
\newblock Spin-glass theory for pedestrians.
\newblock {\em J.STAT.MECH.}, page P05012, 2005.

\bibitem{Dotu}
Iv{\'{a}}n Dot{\'{u}} and Pascal~Van Hentenryck.
\newblock A note on low autocorrelation binary sequences.
\newblock In {\em Principles and Practice of Constraint Programming - {CP}
  2006, 12th International Conference, {CP} 2006, Nantes, France, September
  25-29, 2006, Proceedings}, pages 685--689, 2006.

\bibitem{Gallardo}
Jos{\'{e}}~E. Gallardo, Carlos Cotta, and Antonio~J. Fern{\'{a}}ndez.
\newblock A memetic algorithm for the low autocorrelation binary sequence
  problem.
\newblock In {\em Genetic and Evolutionary Computation Conference, {GECCO}
  2007, Proceedings, London, England, UK, July 7-11, 2007}, pages 1226--1233,
  2007.

\bibitem{Golay}
Marcel J.~E. Golay.
\newblock The merit factor of long low autocorrelation binary sequences.
\newblock {\em {IEEE} Trans. Information Theory}, 28(3):543--549, 1982.

\bibitem{Halim}
Steven Halim, Roland H.~C. Yap, and Felix Halim.
\newblock Engineering stochastic local search for the low autocorrelation
  binary sequence problem.
\newblock In {\em Principles and Practice of Constraint Programming, 14th
  International Conference, {CP} 2008, Sydney, Australia, September 14-18,
  2008. Proceedings}, pages 640--645, 2008.

\bibitem{Kratica12}
Jozef Kratica.
\newblock A mixed integer quadratic programming model for the low
  autocorrelation binary sequence problem.
\newblock {\em Serdica J. Comput.}, 6(4):385--400, 2012.

\bibitem{kschischang2001factor}
Frank~R Kschischang, Brendan~J Frey, and H-A Loeliger.
\newblock Factor graphs and the sum-product algorithm.
\newblock {\em IEEE Transactions on information theory}, 47(2):498--519, 2001.

\bibitem{Marinari94}
Enzo Marinari, Giorgio Parisi, and Felix Ritort.
\newblock Replica field theory for deterministic models. {I}. {B}inary
  sequences with low autocorrelation.
\newblock {\em J. Phys. A}, 27(23):7615--7645, 1994.

\bibitem{Mertens96}
S.~Mertens.
\newblock Exhaustive search for low-autocorrelation binary sequences.
\newblock {\em J. Phys. A}, 29(18):L473--L481, 1996.

\bibitem{dilutedricci}
M.~Mezard, F.~Ricci-Tersenghi, and R.~Zecchina.
\newblock Alternative solutions to diluted p-spin models and xorsat problems.
\newblock {\em J.STAT.PHYS.}, 111:505, 2003.

\bibitem{MPV87}
Marc M\'ezard, Giorgio Parisi, and Miguel~Angel Virasoro.
\newblock {\em Spin glass theory and beyond}, volume~9 of {\em World Scientific
  Lecture Notes in Physics}.
\newblock World Scientific Publishing Co., Inc., Teaneck, NJ, 1987.

\bibitem{SP}
Marc M{\'e}zard and Riccardo Zecchina.
\newblock Random $k$-satisfiability problem: From an analytic solution to an
  efficient algorithm.
\newblock {\em Phys.\ Rev.\ E}, 66:056126, 2002.

\bibitem{Migliorini94}
Gabriele Migliorini and Felix Ritort.
\newblock Dynamical behaviour of low autocorrelation models.
\newblock {\em J. Phys. A}, 27(23):7669--7686, 1994.

\bibitem{Militzer}
Burkhard Militzer, Michele Zamparelli, and Dieter Beule.
\newblock Evolutionary search for low autocorrelated binary sequences.
\newblock {\em {IEEE} Trans. Evolutionary Computation}, 2(1):34--39, 1998.

\bibitem{Mulet02}
Roberto Mulet, Andrea Pagnani, Martin Weigt, and Riccardo Zecchina.
\newblock Coloring random graphs.
\newblock {\em CoRR}, cond-mat/0208460, 2002.

\bibitem{SR14}
B~Suribabu Naick and P~Rajesh Kumar.
\newblock Detection of low auto correlation binary sequences using meta
  heuristic approach.
\newblock {\em International Journal of Computer Applications}, 106(10), 2014.

\bibitem{Mertens16}
Tom Packebusch and Stephan Mertens.
\newblock Low autocorrelation binary sequences.
\newblock {\em J. Phys. A}, 49(16):165001, 18, 2016.

\bibitem{CVM}
Tommaso Rizzo, Alejandro Lage-Castellanos, Roberto Mulet, and Federico
  Ricci-Tersenghi.
\newblock Replica cluster variational method.
\newblock {\em J.\ Stat.\ Phys.}, 139:375, 2010.

\bibitem{Weigt06}
Martin Weigt and Haijun Zhou.
\newblock Message passing for vertex covers.
\newblock {\em Phys. Rev. E}, 74:046110, Oct 2006.

\bibitem{GBP}
Jonathan~S Yedidia, William~T Freeman, and Yair Weiss.
\newblock Understanding belief propagation and its generalizations.
\newblock {\em Exploring artificial intelligence in the new millennium},
  8:236--239, 2003.

\bibitem{Yedidia02}
Jonathan~S. Yedidia, William~T. Freeman, and Yair Weiss.
\newblock Constructing free-energy approximations and generalized belief
  propagation algorithms.
\newblock {\em {IEEE} Trans. Information Theory}, 51(7):2282--2312, 2005.

\bibitem{ZPhD08}
L.~{Zdeborov{\'a}}.
\newblock {\em {Statistical Physics of Hard Optimization Problems}}.
\newblock PhD thesis, PhD Thesis, 2008, June 2008.

\end{thebibliography}

\end{document}